\documentclass[aps,prl,reprint,superscriptaddress]{revtex4-2}

\usepackage{textcomp}
\usepackage{graphicx}
\usepackage{amsmath}
\usepackage{amssymb}
\usepackage{siunitx}
\usepackage[colorlinks=true, citecolor={black}, urlcolor={black}, linkcolor = {black}]{hyperref}
\usepackage{cleveref}
\Crefname{figure}{Fig.}{Figs.}
\crefname{figure}{Fig.}{Figs.}

\makeatletter

\newcommand{\fmarki}{*}
\newcommand{\fmarkii}{\ensuremath{\dagger}}
\newcommand{\fmarkiii}{\ensuremath{\ddagger}}
\newcommand{\fmarkiv}{\ensuremath{\mathsection}}
\newcommand{\fmarkv}{\ensuremath{\mathparagraph}}
\newcommand{\fmarkvi}{\ensuremath{\|}}
\newcommand{\fmarkvii}{**}
\newcommand{\fmarkviii}{\ensuremath{\dagger\dagger}}
\newcommand{\fmarkix}{\ensuremath{\ddagger\ddagger}}
\def\@fnsymbol#1{{\ifcase#1\or \fmarki\or \fmarkii\or \fmarkiii\or \fmarkiv\or \fmarkv\or \fmarkvi\or \fmarkvii\or \fmarkviii\or \fmarkix \else\@ctrerr\fi}}
\makeatother

\renewcommand{\fmarki}{\ensuremath{\dagger}}
\renewcommand{\fmarkii}{*}

\def\uppi~{$\mathrm{\pi}$}

\newcommand{\var}[2]{$#1_{\mathrm{#2}}$}

\newcommand{\ABS}[1]{$V_{\mathrm{ABS}}^{\mathrm{(#1)}}$}

\begin{document}

\title{Using Andreev bound states and spin to remove domain walls in a Kitaev chain}

\author{Wietze D. Huisman}
\altaffiliation{These authors contributed equally to this work.}
\affiliation{QuTech and Kavli Institute of Nanoscience, Delft University of Technology, Delft, 2600 GA, The Netherlands}

\author{Sebastiaan L. D. ten Haaf}
\altaffiliation{These authors contributed equally to this work.}
\affiliation{QuTech and Kavli Institute of Nanoscience, Delft University of Technology, Delft, 2600 GA, The Netherlands}

\author{Chun-Xiao Liu}
\affiliation{QuTech and Kavli Institute of Nanoscience, Delft University of Technology, Delft, 2600 GA, The Netherlands}

\author{Qingzhen Wang}
\affiliation{QuTech and Kavli Institute of Nanoscience, Delft University of Technology, Delft, 2600 GA, The Netherlands}
\author{Alberto Bordin}
\affiliation{QuTech and Kavli Institute of Nanoscience, Delft University of Technology, Delft, 2600 GA, The Netherlands}
\author{Florian J. Bennebroek Evertsz'}
\affiliation{QuTech and Kavli Institute of Nanoscience, Delft University of Technology, Delft, 2600 GA, The Netherlands}
\author{Bart Roovers}
\affiliation{QuTech and Kavli Institute of Nanoscience, Delft University of Technology, Delft, 2600 GA, The Netherlands}
\author{Michael Wimmer}
\affiliation{QuTech and Kavli Institute of Nanoscience, Delft University of Technology, Delft, 2600 GA, The Netherlands}

\author{Srijit Goswami}\email{s.goswami@tudelft.nl}
\affiliation{QuTech and Kavli Institute of Nanoscience, Delft University of Technology, Delft, 2600 GA, The Netherlands}


\begin{abstract}
Quantum dot-superconductor hybrids have been established as a suitable platform for realizing Kitaev chains hosting Majorana bound states.
Implementing these structures in a qubit architecture is expected to result in coherence times that scale exponentially with the lengths of the chains.
To scale to longer systems, the phase differences between all superconducting segments in the chain need to be controlled.
While this control has been demonstrated by using an external magnetic flux~\cite{tenHaaf2025,bordin2025_qdprobe}, ideally it can be achieved with control over intrinsic system parameters~\cite{CX_scaling}.
In this work, we investigate whether the relevant phase differences can be tuned through the spin degree of freedom in each QD, or the chemical potential of the discrete bound states in the hybrid sections. 
We confirm that both these tuning knobs allow for controlling the phase difference in the couplings between neighbouring QDs, bypassing the requirement to tune an external flux.
However, we find that the amplitude of the phase shifts can deviate from a discrete $\pi$-shift. 
We introduce a spatial variation in the spin-orbit field as a possible mechanism to explain the observed behaviour and comment on the consequences for experimentally creating long Kitaev chains.
\end{abstract}

\maketitle
\section{Introduction}
The Kitaev chain models a 1D spinless fermionic chain with a p-wave superconducting pairing, which hosts unpaired Majoranas at its edges~\cite{Kitaev2001}.
These quasiparticles are central to many schemes for fault-tolerant quantum computation~\cite{Kitaev2003,Nayak2008}.
Semiconductor quantum dots coupled via superconductors offer an experimentally tunable realization of the Kitaev chain~\cite{DasSarma2012,Leijnse2012,Fulga2013,Tsintzis2022}, which has so-far enabled the creation of localized Majoranas in systems with two~\cite{Dvir2023,tenHaaf2024,Zatelli2024} and three quantum dot (QD) sites~\cite{Bordin2025,tenHaaf2025}.
The protection of Majoranas against perturbations scales with the number of sites in the chain, making it highly desirable to extend these arrays.
Scaling to longer chains would, for instance, enable implementations of a parity qubit with longer coherence times~\cite{Leijnse2012,Tsintzis2024_roadmap,Bordin2025} or demonstrations of the non-abelian exchange of Majorana quasiparticles~\cite{Svend2022,Boross2024_braiding, Tsintzis2024_roadmap,miles2025_braiding}.
A fundamental obstacle for scaling to longer chains is posed by phase differences that arise between neighbouring superconductors.
In a Kitaev chain, any $\pi$-phase difference between neighbouring superconducting pairings locally closes the excitation gap.
This effectively cuts the chain in segments as a so-called domain wall is formed~\cite{DasSarma2012,Kitaev2001}.
One ideally requires all phases to be uniform.
Recent works on three-site systems address this problem via a flux-controlled superconducting loop~\cite{tenHaaf2025, bordin2025_qdprobe}.
This is, however, not a practical solution for longer systems: an N-site chain would require independent control over the flux through N-2 superconducting loops.\newline

It was recently proposed~\cite{CX_scaling} that  Kitaev chains could in fact be scaled without explicit flux control.  
Instead, control over the spin configuration of the QDs and/or control over the chemical-potential energy of the hybrid segments was predicted to be sufficient to achieve this.
Here, we investigate this strategy experimentally, by studying a three-site Kitaev chain in an InSbAs 2DEG \textit{with} a flux-tunable superconducting loop. 
We follow the protocol in Ref.~\cite{tenHaaf2025} to create localized Majorana edge modes, and  study the flux-response of the excitation gap in the middle quantum dot. 
By comparing the response to the superconducting flux for different spin configurations, we confirm that changing the spin polarization of one of the quantum dots introduces a discrete phase shift in the system.
Furthermore, we show that changing the chemical potential energy of one of the hybrid segments induces a similar phase shift, providing a second mechanism that can remove a domain wall.
Closer inspection reveals that the phase shifts evolve smoothly as a function of the ABS charge, which can result in sweet spots with an arbitrary (non-$\pi$) difference in their relative phase. 
We theoretically analyse how an effective spatial variation in the spin-orbit field gives rise to the smooth phase evolution and derive an analytical expression to compare to the experimental results.
Our results show that flux-free scaling of the Kitaev chain can be feasible, but that a deeper understanding of microscopic details of the system is important for implementing the technique.
\subsection*{The scaling protocol}
\begin{figure}[t!]
\centering
    \includegraphics[width = 0.5\textwidth]{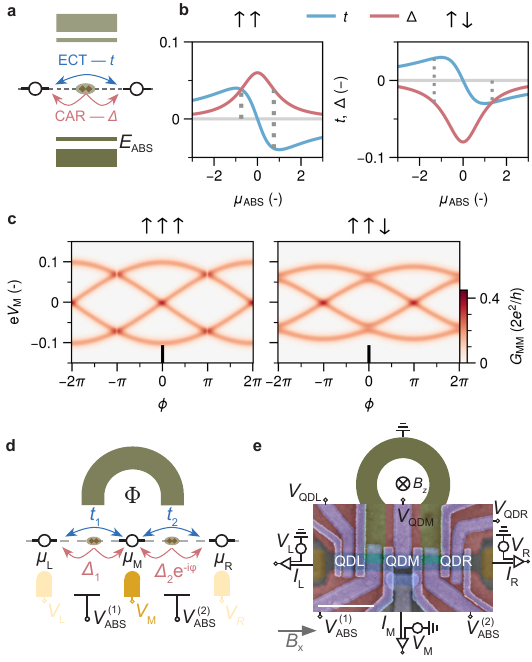}
    \caption{\textbf{Proposed mechanisms for controlling phase in the Kitaev chain and device layout of a three-site chain.} \textbf{a)} Schematic of ECT and CAR sub-gap processes in a two-site Kitaev chain setup, controlled by an ABS with energy $E_{\mathrm{ABS}}$. \textbf{b)} ECT ($t$) and CAR ($\Delta$) amplitudes as a function of \var{\mu}{ABS} for two different spin configurations. 
    \textbf{c)} Simulated conductance spectra on the middle quantum dot of a three-site Kitaev chain, as a function of the phase $\phi$. 
    Two different spin configurations are compared, demonstrating the removal of a domain wall through flipping the rightmost spin. 
    \textbf{d)} Three-site Kitaev chain schematic showing the two pairs of couplings $t_1,\Delta_1$ and $t_2,\Delta_2$. 
    Phase modulation is enabled by threading a flux $\Phi$ through the superconducting loop connecting the two hybrid sections. 
    \textbf{e)} SEM of the measured device. An out-of plane magnetic field $B_z$ is used to modulate the flux, with a period of \SI{28}{\micro\tesla}. An in-plane magnetic field $B_x$ is applied parallel to the channel, to spin-polarize the quantum dots.  
    }
    \label{fig:Fig1}
\end{figure}

The Hamiltonian for the general N-site Kitaev chain is given by~\cite{Kitaev2001,DasSarma2012}:
\begin{equation*}
    H = \sum_j^N\mu_jc^{\dagger}_jc_j+\sum_j^{N-1}\left(t_j c^{\dagger}_{j}c_{j+1} +\lvert\Delta_j\rvert e^{-i\phi_j}c^{\dagger}_jc^{\dagger}_{j+1}\right)+h.c.
\end{equation*}
A gauge transformation allowed for taking the chemical potentials $\mu_j$ and hopping terms $t_j$ to be real and assigning a complex phase $\phi_j$ only to the superconducting pairings $\Delta_j$~\cite{Kitaev2001,Alicea_2012}.
In a 1D system with Rashba spin-orbit coupling, these phases are restricted so that $\phi_j$ is either 0 or $\pi$~\cite{DasSarma2012}.
For any $N$, unpaired Majoranas arise on the two outermost sites when $\mu_j=0$, $\lvert{t_j}\rvert=\lvert\Delta_j\rvert$, this is referred to as the Majorana sweet spot.
When all phases are equal (such that either all $t_j=+\Delta_j$ or all $t_j=-\Delta_j$), the middle sites host an excitation gap that separates the Majorana modes at the edge.\newline\newline
The Kitaev chain can be effectively engineered by coupling spin-polarized QDs via Andreev bound states in a semiconductor–superconductor hybrid segment~\cite{Tsintzis2024_roadmap, Dvir2023, tenHaaf2024}.
Focusing first on a minimal two-site system, the hopping interaction ($t$) occurs through elastic co-tunnelling (ECT), whereas crossed Andreev reflection (CAR) provides a pairing interaction ($\Delta$) via the creation or splitting of Cooper pairs in the superconductor~\cite{Liu2022_carect,Bordin2023_carect} (\cref{fig:Fig1}a).
Crucially, the amplitudes $t$, $\Delta$ of these virtual processes depend differently on the ABS chemical potential energy $\mu_{\mathrm{ABS}}$.
Additionally, the relative signs of the amplitudes are affected by the spin configuration of the spin-polarized quantum dots. 
Depending on whether the spins of the QDs are aligned or anti-aligned, either the CAR or the ECT process requires the presence of spin-orbit interactions. 
As a result, the coupling mechanism differs for different spin configurations. 
\Cref{fig:Fig1}b shows theoretically calculated $t,\Delta$ amplitudes as a function of $\mu_{\mathrm{ABS}}$, previously shown in Ref.~\cite{CX_scaling}, comparing an equal spin alignment and an opposite spin-alignment in the QDs. 
For each spin configuration, tuning $\mu_{\mathrm{ABS}}$ enables either $t=+\Delta$ or $t=-\Delta$.
Focussing on $\mu_{\mathrm{ABS}}<0$, changing the spin of one of the QDs similarly changes the relative amplitude from $t = +\Delta$ to $t = -\Delta$ (vice versa for $\mu_{\mathrm{ABS}}>0$). \newline\newline
The relative sign between $t,\Delta$ becomes relevant when considering a Kitaev chain with $N\geq3$ QD sites.
To see this, we turn to a spinful model considering three spinful QDs and two ABSs (see Methods).
\Cref{fig:Fig1}c shows numerical conductance spectra of the middle quantum dot at a three-site sweet spot, as a function of the phase difference $\phi$ between the two superconducting pairings.
For the left spin configuration, for example, we find a situation where the gap is closed when $\phi=0$, due to the formation of a domain wall.
Changing the spin on the right QD shifts the entire spectrum by a $\pi$-phase, removing the domain wall and reopening the excitation gap in the middle QD at $\phi=0$.
\subsection*{Device and measurement set-up}
A schematic overview and a Scanning Electron Micrograph (SEM) of the measured device are shown in \cref{fig:Fig1}d and \cref{fig:Fig1}e. 
Three large lithographically defined depletion gates (red) confine a narrow conductance channel across two superconducting strips (green).
Narrow gates are used to define three quantum dots on the left, in the middle and on the right of the two superconducting regions.
Their chemical potential energies are controlled by \var{V}{QDL}, \var{V}{QDM} and \var{V}{QDR} respectively. 
The two hybrid sections host discrete Andreev bound states~\cite{tenHaaf2025}, whose energies are controlled by \ABS{1} and \ABS{2}. 
The sections are connected in a loop, enabling an out-of-plane magnetic field $B_z$ to control the flux $\Phi$ through the loop.
This directly controls the relative phases of the inter-dot couplings between neighbouring QD pairs~\cite{tenHaaf2025, bordin2025_qdprobe}.
Each QD is coupled to a normal contact via a tunnelling barrier, allowing us to probe the local density of states at the edges and in the middle of the chain. 
Fast measurements are performed using radio-frequency lead reflectometry techniques enabled by off-chip lumped-element resonators, where the reflected phase and amplitude of signal $S_{21}^i$ on lead $i$ is recorded.
We plot the processed signal $\widetilde{S}^i_{21}$, as detailed in Methods.
In addition, we show measurements of the differential conductance $G_{\mathrm{MM}}=\frac{dI_{\mathrm{M}}}{dV_{\mathrm{M}}}$ through the middle ohmic contact.

\section*{Results}
\subsection{Spin-induced phase shift}
\begin{figure}
\centering
    \includegraphics[width = 0.5\textwidth]{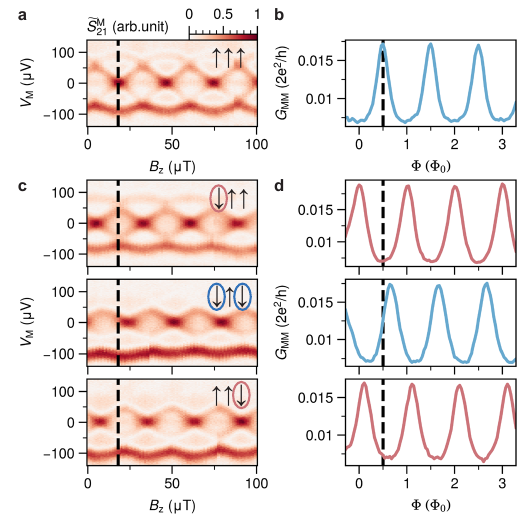}
    \caption{\textbf{Spin-induced phase shifts.} \textbf{a)} Bias spectroscopy measurement for the middle quantum dot as a function of the out-of-plane magnetic field \var{B}{z} for an $\uparrow\uparrow\uparrow$ spin configuration. \textbf{b)} A higher resolution \var{G}{MM} linetrace along \var{V}{M}~=~0, expressed in units of the flux quantum \var{\Phi}{0}. 
    We identify a flux where the excitation gap closes.
    \textbf{c)} Repeated measurements of a), for different spin-configurations.
    \textbf{d)} Comparison of \var{G}{MM} linetraces along \var{V}{M}=0, for the same spin-configurations as in c).
    Applied magnetic field along \var{B}{x} is \SI{150}{\milli\tesla}.
    The behaviour is reproducibly observed, shown in \cref{fig:supp2}.
    }
    \label{fig:Fig2}
\end{figure}

To initialize the system, we identify an isolated orbital in each of the three QDs.
An in-plane magnetic field (\var{B}{x}) is applied along the length of the 1D channel, to spin polarize the QDs (\cref{fig:supp1}).
Since each orbital has two spin polarizations, we can study eight different combinations of resonances. 
We use the combined spin-configuration as label.
Following the same protocol as in our previous work~\cite{tenHaaf2025}, the device is first tuned up so that the condition $\lvert{t_i}\rvert=\lvert\Delta_i\rvert$ is satisfied for each of the two QD pairs (left/middle and middle/right).
This is achieved by tuning \ABS{1} and \ABS{2} to change the charge and energy of ABSs, which in turn affects the interdot couplings (see \cref{fig:Fig1}b,c).
The tune-up is completed by setting all \var{V}{QDi} to the charge degeneracy point corresponding to $\mu_i=0$.
To identify the flux corresponding to a phase $\phi=0$, we measure the finite-bias conductance through the middle quantum dot \var{G}{MM} as a function of \var{B}{z} (\cref{fig:Fig2}a).
The excitation gap in the middle QD closes periodically as a function of $B_{z}$, in line with the predicted behaviour shown in \cref{fig:Fig1}c and Refs.~\cite{CX_scaling,tenHaaf2025, bordin2025_qdprobe}.
The value of \var{B}{z} where the excitation gap in the middle QD closes is taken as reference point (indicating $\phi$=$\pi$), which can be easily extracted from a measurement along \var{V}{M}~=~0 (\cref{fig:Fig2}b).\\ \\
The result in \cref{fig:Fig2}a is obtained for an $\uparrow\uparrow\uparrow$ spin configuration.
We can select a different spin configuration through the voltages $V_{\mathrm{QD}i}$.
For each adjustment, a slight retuning of $V_{\mathrm{ABS}}$ is required to maintain the $|t_i| = |\Delta_i|$ condition. 
We repeat the measurement in \cref{fig:Fig2}a for three different combinations of spin polarizations of the outer QDs, shown in \cref{fig:Fig2}c.
For every adjustment of a spin polarization, we observe that the spectrum shifts by approximately half a flux period, corresponding to a $\pi$-shift in the effective phase $\phi$.
To compare the flux response more closely, we compare linetraces along \var{V}{M} = 0 in \cref{fig:Fig2}d.
Focusing on the $\Phi$ value indicated by the dashed line, at fixed \var{B}{x}, the excitation gap in the system is either present or closed depending on the spin-configuration.
This in principle demonstrates the desired control: the spin configuration serves as a tool to remove a domain wall, without requiring to tune the external flux.
We note, however, that small offsets from $\phi=0$ or $\phi=\pi$ appear in the bottom two configurations, marking a departure from the discrete 0/$\pi$-shift presented in \cref{fig:Fig1}c. 
This motivates a more in-depth study where we turn to the second predicted control knob: the Andreev bound state charge.
\subsection*{ABS-induced phase shift}
\begin{figure}
\centering
    \includegraphics[width = 0.5\textwidth]{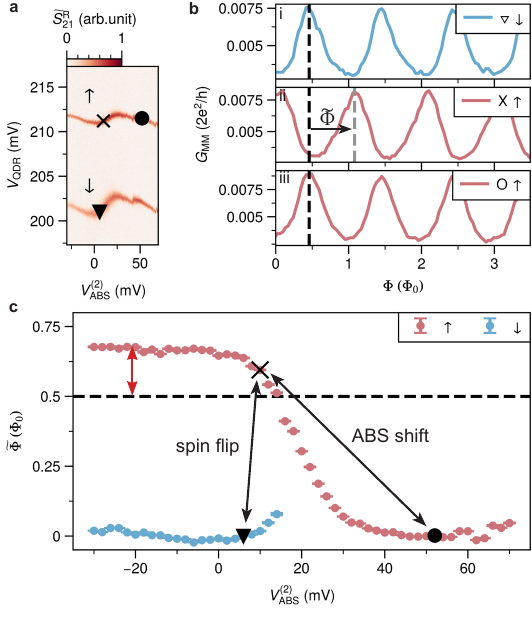}
    \caption{\textbf{ABS-induced phase shifts.} \textbf{a)} Hybridization between the right ABS and two spin levels of the right quantum dot QDR. Three sweet spots ($\times,\circ,\triangledown$) are identified from charge stability diagrams of the right and middle quantum dot (\cref{fig:supp4}). 
    \textbf{b)} Flux dependence of the zero-bias conductance \var{G}{MM} on the middle dot, measured at the three sweet spots indicated in a). 
    \textbf{c)} Continuous measurement of the extracted flux-shift $\widetilde \Phi$ as a function of the right ABS energy. 
    The extracted shift $\widetilde \Phi$ is defined with respect to the first conductance peak in the zero-bias trace as indicated in b) (see \cref{fig:supp3} for further details about the extraction). 
    The spin-down configuration is only shown for values of \ABS{2} $<$ \SI{16}{\milli\volt}, as the extraction was unreliable for the rest of the range.
    Raw datasets are shown in \cref{fig:supp6}, \ref{fig:supp7}.
    Measurements were performed at an in-plane field $B_{\mathrm{x}}$ of \SI{200}{\milli\tesla}.
    The flux dependence of the energy spectrum at all values of the ABS energy is shown in \cref{fig:supp5}.}
    \label{fig:Fig3}
\end{figure}

The charge of the Andreev bound states provides a second way to change the relative sign of the $t$ and $\Delta$ amplitudes, as highlighted in \cref{fig:Fig1}b.
Hence, tuning between the two available $|t|=|\Delta|$ sweet-spots within the range of a single ABS should yield a relative phase shift, while maintaining the same spin-configuration. 
In contrast to changing the spin of the QDs, this control via \var{V}{ABS} allows for a continuous interpolation between two sweet spots.
First, we search for a range in \ABS{2} where the rightmost QD interacts with a single ABS.
\cref{fig:Fig3}a shows a charge stability diagram of two spin-polarized resonances in the right QD interacting with an ABS in the right hybrid segment.
As \ABS{2} is varied, the positions of the QD resonances are modulated, indicating interaction with ABSs in this range~\cite{Rasmussen2018}. 
Here, we find that one sweet spot can be clearly identified for the spin-down resonance, while the spin-up resonance allows for determining a sweet spot at two different ABS energies (\cref{fig:supp4}). 
At each of these three sweet spots, we obtain a zero-bias conductance measurement as a function of flux, shown in \cref{fig:Fig3}b.  
Comparing these linetraces, we observe that now both the change in spin and the change of ABS energy result in comparable relative phase shifts. 
Next, we study the flux dependence of the system throughout a larger range of \ABS{2} around the two spin-up sweet spots, to interpolate how the system evolves from the result in panel \ref{fig:Fig3}b.ii to the shifted trace in panel \ref{fig:Fig3}b.iii.
Generally then $\lvert{t_2}\rvert\neq\lvert\Delta_2\rvert$, while we maintain $\lvert{t_1}\rvert=\lvert\Delta_1\rvert$.
Similar to \cref{fig:Fig3}b, we characterise the flux dependence by measuring the response of the zero-bias conductance. 
The oscillations are fitted to estimate the value \var{B}{z} where the excitation gap closes (see \cref{fig:supp3}), which allows to extract a single value $\widetilde \Phi$ that characterises the relative flux-shift at each \ABS{2} point with respect to the dashed line in panel 3b.i.
The evolution of this extracted flux-shift $\widetilde \Phi$ versus the ABS energy is shown in \cref{fig:Fig3}c.\newline\newline
The extracted flux-shift in \cref{fig:Fig3}c reveals three important observations. 
First, we find that the flux-shift induced by a spin-flip is similar to the flux-shift acquired by changing the ABS charge from one sweet spot to another.
This indicates that the underlying mechanism shown in \cref{fig:Fig1}b appears to be well-understood. 
However, we also observe that the flux-shift, and therefore the phase, varies smoothly between two-sweet spots as the ABS energy is changed, deviating from the scenario where only phases of $0$ or $\pi$ appear in the system~\cite{DasSarma2012}.
Given that the exact positions of the two sweet spots along \var{V}{ABS} depend on underlying microscopic details, a smooth phase evolution means that any arbitrary flux-shift can appear in practice.
Furthermore, the phase separation of the plateaus at the extremes of the ABS range is larger than $\pi$, indicated by the red arrow.
The practical implications of this are treated in the Discussion.
Below we address how the model underlying \cref{fig:Fig1}c can be adjusted to reproduce this behaviour. 
 
\subsection*{Possible origin of the smooth phase evolution}
\begin{figure}
\centering
\includegraphics[width = 0.5\textwidth]{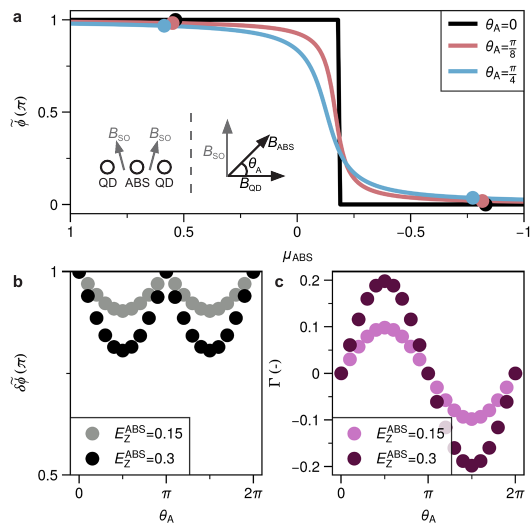}    \caption{\textbf{Theoretical analysis of $\widetilde{\phi}$ as a function of $\mathbf{\mu_{\mathrm{ABS}}}$.} \textbf{a)} To explain the experimental result, we propose a spatial variation in the spin-orbit hopping between neighbouring sites as possible source (inset left). 
We implement this here equivalently through an offset angle $\theta_{\mathrm{A}}$ between the polarization of the QDs and the ABS (inset right).
The more general case is discussed in Methods.
The analytically obtained phase-shift $\widetilde\phi$ (\cref{eq:phasedependence}) is plotted, for various $\theta_{\mathrm{A}}$. 
For each curve, the two $\mu_{\mathrm{ABS}}$ values are indicated that correspond to a sweet-spot numerically.
A fitting of the experimental data to the theoretical result is shown in \cref{fig:supp5}.
\textbf{b)} The smoothening of the phase dependence on $\mu_{\mathrm{ABS}}$ results in a decrease in the phase difference $\delta\widetilde{\phi}$ between two sweet-spots, for two values of $E^{\mathrm{ABS}}_Z$ in units of the induced superconducting pairing in the ABS.
\textbf{c)} The smoothening parameter $\Gamma$ as a function of $\theta_{\mathrm{A}}$, for two values of $E^{\mathrm{ABS}}_Z$. 
}\label{fig:Fig4}
\end{figure}

The above results deviate from the original theoretical result~\cite{CX_scaling} in two clear ways: the system \textit{smoothly} interpolates between two phase plateaus as a function of \ABS{2}, and the separation of the two plateaus corresponds to a total phase-shift larger than $\pi$.
In a system that respects complex conjugation symmetry, only discrete 0 or $\pi$ phases arise. 
A possible mechanism that could break this symmetry, is a spatial variation in the effective spin-orbit field along the chain, which was so far assumed to be fixed perpendicular to the chain.
As a starting point, we present here a simple adjustment to the model underlying \cref{fig:Fig1}c that implements this by introducing an in-plane offset angle $\theta_{\mathrm{A}}$ between the polarization-axis of the QDs ($\vec{B}_{\mathrm{QD}}$) and that of the ABS ($\vec{B}_{\mathrm{ABS}}$), with Zeeman energy $E^{\mathrm{ABS}}_{\mathrm{Z}}$.
This is equivalent to a specific configuration of a spatially varying spin-orbit field $\vec{B}_{\mathrm{SO}}$,  treated more generally in Methods.
Mapping the full model onto an effective three-site model allows to obtain an expression for the complex phase-shift $\widetilde\phi$ in the system (see Methods) as a function of ABS parameters:
\begin{equation}  
\label{eq:phasedependence}
\widetilde{\phi}=\frac{\pi}{2}-\arctan\left[\frac{\mu_{\mathrm{ABS}}+\mu_0(\theta_{\mathrm{A}},E_{\mathrm{Z}}^{\mathrm{ABS}})}{\Gamma(\theta_{\mathrm{A}},E^{\mathrm{ABS}}_{\mathrm{Z}})}\right]
\end{equation}
When $\theta_{\mathrm{A}},E^{\mathrm{ABS}}_{\mathrm{Z}}=0$, the dependence on $\mu_{\mathrm{ABS}}$ is a step-function between 0 and $\pi$, centred around $\mu_{\mathrm{ABS}}=0$.
The parameter $\Gamma\propto E^{\mathrm{ABS}}_{\mathrm{Z}}\sin\left(\theta_{\mathrm{A}}\right)$ smoothens the step-function and $\mu_0\propto E^{\mathrm{ABS}}_{\mathrm{Z}}\cos(\theta_{\mathrm{A}})$ shifts the centre of the curve away from $\mu_{\mathrm{ABS}}=0$.
In \cref{fig:Fig4}a, we show the dependence on $\mu_{\mathrm{ABS}}$ for varying $\theta_{\mathrm{A}}$ at fixed $E^{\mathrm{ABS}}_{\mathrm{Z}}$, highlighting the smoothening effect on the step-function when $\theta_{\mathrm{A}}\neq0$.
The scattered points indicate the two values of $\mu_{\mathrm{ABS}}$ that numerically correspond to sweet spots. 
Due to the smoothening, the phase separation between the two sweet-spots decreases as a function of $\theta_{\mathrm{A}}$ and $E^{\mathrm{ABS}}_{\mathrm{Z}}$ (\cref{fig:Fig4}b), as they lie more towards the centre of the smoothened curve, similar to \cref{fig:Fig3}c.
A plot of $\Gamma$ as a function of $\theta_{\mathrm{A}}$ is shown for two values of $E^{\mathrm{ABS}}_{\mathrm{Z}}$ in \cref{fig:Fig4}c, where we see that the broadening is maximum when $\theta_{\mathrm{A}}=\pi/2$.
Fitting the result \cref{eq:phasedependence} to the experimental data captures the overall trend well~(\cref{fig:supp5}).
We stress, however, that the larger-than-pi separation of the plateaus in \cref{fig:Fig3}c is not directly described by this mechanism.

\section{Discussion}
The key finding of this work is that both quantum dot spin and ABS charge \textit{can} remove a domain wall in a Kitaev chain device.
However, the resulting phase shifts may differ from the ideal value of $\pi$.
In our model, we find that spatial variations in the alignment of the spin-orbit field can be a possible explanation for these deviations.
We believe this to be a physically reasonable assumption, which may for example arise through variations in the electrical fields or microscopic differences in the QDs/tunnel barriers.
An anisotropic g-tensor along the chain could also produce this, although in this specific material the in-plane g-factor is expected to be isotropic~\cite{Metti2023}.
The larger-than-$\pi$ variation may arise if the spin-orbit field itself is affected by the \var{V}{ABS} gate voltage, or through the presence of multiple Andreev bound states.
To minimize the deviations induced by spin-orbit misalignments, this effect should be taken into account in the design of future experiments. 
Specifically, our results suggest that complex geometries for applications such as braiding experiments should consist of predominantly parallel chains, allowing a single field $\vec{B}$ to be perpendicular to $\vec{B}_{\mathrm{SO}}$ everywhere.
\newline \newline
We emphasize that further work is needed for a complete understanding of the exact underlying mechanism.
We expect, for instance, that an effective change in the offset angle $\theta_{\mathrm{A}}$ may be observed when using different ABS orbitals as mediator of the interdot coupling.
Further, we find experimentally that $\widetilde{\Phi}$ is affected by the direction of the external in-plane magnetic field used to spin-polarize the QDs, shown in \cref{fig:supp9}.
Importantly, phase shifts extending beyond half a flux period become visible only when tracking the phase dependence as a function of a continuous parameter, most clearly seen in the raw data shown in \cref{fig:supp5}.
For the data in \Cref{fig:Fig2} on the other hand, whether the phase-shifts are smaller or larger than $\pi$ remains ambiguous.
Ref.~\cite{bordin2025_qdprobe} reports similar deviations from $\pi$-shifts in a similar device, but can not resolve the exact amplitude for the same reason.
We hope further experiments can resolve the outstanding questions.
\newline\newline
The main purpose of the investigated protocol is to remove domain walls in long Kitaev chains and maximize the bulk excitation gap, without requiring an external flux.
The main consequence of the observed deviations from $\pi$-shifts, is that the technique can lead to a reduced excitation gap on the middle site compared to the maximum when the effective phase $\phi=0$.
We show this reduction explicitly in \cref{fig:supp8}.
As an example, we can quantitatively analyse the reduction in the excitation gap for the results in \cref{fig:Fig2}.
By extracting the phases of the $\downarrow\uparrow\uparrow$ and $\downarrow\uparrow\downarrow$ configurations in \cref{fig:Fig2}, we find that the phase difference between the states deviates from $\pi$ by approximately $30\%$, or \SI{4}{\micro\tesla} (\cref{fig:supp10}). 
This in turn leads to a decrease in the excitation gap from approximately \SI{50}{\micro\electronvolt} to \SI{38}{\micro\electronvolt}.
A larger deviation from $\pi$ translates directly to a larger reduction in the energy gap. 
However, as long as the phase shift is non-zero, it may be used to avoid a situation where the excitation gap closes.  
Compared to fine-tuned flux control over multiple loops, with the drawbacks of requiring additional control lines and accounting for cross-talk~\cite{Prosko2024, Lykkegaard2025}, we believe this presented technique to be the preferred alternative.

\section*{Conclusions}
We have studied the phase-dependent energy spectra of different spin-configurations of a three-site Kitaev chain. 
Through flux-dependent measurements, we achieve phase shifts close to $\pi$ by changing the spin occupation on either of the outer QDs. 
The phase shifts observed in this work would allow to avoid domain walls, enabling the scaling to longer Kitaev chains without the need for external flux control. 
Interestingly, upon further investigation we find that the phase shift varies smoothly with the ABS energy, which can give rise to arbitrary phase shifts between sweet-spots.
We theoretically show that a spatial variation in the spin-orbit field can be a possible mechanism for the observed continuous transition. 
This finding helps to understand the origin of phase shifts that deviate from $\pi$, and helps to understand conditions under which the spin-flip protocol can be used reliably in future experiments.

\section{Acknowledgements}
We thank C. Thomas, D. Xiao and M. J. Manfra for the provision of the 2DEG materials. 
We thank O.W.B.~Benningshof, T.~Orton and J.D.~Mensingh for technical assistance with the cryogenic electronics. We thank S. Miles and J. L. Torres for their input on the theoretical analysis. We are grateful to I. Kulesh, C. G. Prosko and Y. Zhang for developing the resonator circuits and high-frequency set-up. The research at Delft was supported by the Dutch National Science Foundation (NWO),  Microsoft Corporation Station Q and a grant from Top consortium for Knowledge and Innovation program (TKI). 
S.G. and M.W acknowledge financial support from the Horizon Europe Framework Program of the European Commission through the European Innovation Council Pathfinder grant no. 101115315 (QuKiT).

\section{Author contributions}
Q.W. fabricated the device.
Measurements were performed by W.D.H. and S.L.D.tH. 
Theoretical analysis was provided by C-X.L., S.L.D.tH. and M.W. 
A.B., F.J.B.E. and B.R. helped interpret the experiments to understand the phase behaviour.
The manuscript was written by W.D.H. and S.L.D.tH. with input from all co-authors. 
S.G. supervised the experimental work in Delft.

\section{Data availability}
All raw data obtained in relation to this manuscript and the scripts to produce the figures are available on Zenodo~\cite{ZenodoDataRepo}.

\bibliography{ref} 
\begin{widetext}
\clearpage
\newcommand{\beginsupplement}{%
	\setcounter{section}{0}
	\renewcommand{\thesection}{\arabic{section}}%
	\setcounter{table}{0}
	\renewcommand{\thetable}{S\arabic{table}}%
	\setcounter{figure}{0}
	\renewcommand{\thefigure}
    {S\arabic{figure}}%
	\renewcommand{\theHfigure}{S\arabic{figure}}%
	\setcounter{equation}{0}
	\renewcommand{\theequation}{S\arabic{equation}}%
    \renewcommand{\theHequation}{S\arabic{equation}}%

}
\setcounter{subsection}{1}
\beginsupplement
\section{Methods}
\subsection{Device fabrication}
The device investigated in the main text was fabricated using the techniques described in detail in \cite{Moehle2021} and \cite{WangQThesis}. 
The Aluminium loop structure is defined in an InSbAs-Al chip by wet etching, followed by the deposition of three ohmic Ti/Pd contacts. 
After deposition of \SI{20}{nm} AlOx via \SI{40}{\celsius} atomic layer deposition (ALD), three large Ti/Pd depletion gates are evaporated: one large top depletion gate and two bottom depletion gates each extending halfway.
The bottom consists of two gates in order to independently form the left and right halves of the channel, separated by a thin channel where the middle lead is placed.
The channel width is designed to be \SI{200}{\nano\meter}.
Following a second ALD layer (\SI{20}{\nano\meter} AlOx), a first layer of six Ti/Pd finger gates is evaporated.
These are used for controlling the electrochemical potential energies of the QDs and hybrid regions and for defining the tunneling barrier for the middle contact. 
A third ALD layer (\SI{20}{\nano\meter} AlOx) is deposited, followed by evaporation of the remaining six Ti/Pd finger gates that define the three QDs.
For RF-measurements, superconducting LC-resonator circuits are fabricated on a separate chip with a silicon substrate by etching NbTiN.
To apply DC voltages, bias tees are created by depositing \SI{20}{\nano\meter} Cr structures with resistances of $\approx$~\SI{5}{\kilo\ohm}.
\\ \\
The measured device was cooled down a total of twelve times over the course of 1.5 years. 
The initial cool-downs focussed on the results presented in Refs.~\cite{tenHaaf2025} and \cite{Kulesh2025}. 
The results presented in \cref{fig:Fig2} and \cref{fig:Fig3} were obtained in the seventh and twelfth cooldown, where we focussed systematically on the phase-shifts in the system.
The effects of changing spin or ABS charge to shift the phase was observed in earlier cool-downs, but not addressed systematically in those measurements.
A summary of all observed phase shift data, including data from the earlier cool-downs four and six, is shown in \cref{fig:supp2}.\\ \\
\subsection{DC transport measurements}
Measurements are performed in a dilution refrigerator with a base temperature of \SI{20}{\milli\kelvin}.
We expect the effective electron temperature to be 50-\SI{100}{\milli\kelvin}.
Transport measurements presented in the main text are performed in AC and DC using a four-terminal set-up (three ohmic contacts plus two aluminium strips connected in a loop).
The aluminium strips induce a gap of $\approx$~\SI{220}{\micro\volt} \cite{tenHaaf2025} and are kept electrically grounded. 
Each ohmic lead is connected to a current meter and biased through a digital-to-analogue converter and both DC and AC voltages can be applied.
Offsets of the applied voltage-bias on each lead are corrected via independently measuring the Coulomb peaks in the QDs and looking at the change in sign of the current.
The voltage outputs of the current meters are recorded with three digital multimeters and three lock-in amplifiers.  
When applying a DC voltage to one lead (e.g., \var{V}{L}),  the other leads (i.e., \var{V}{M} and \var{V}{R}) are kept grounded.
AC excitations are applied with amplitudes around \SI{5}{\micro \volt} RMS and a frequency of \SI{23}{\hertz}.
In this way, the conductance of the middle lead $G_{\mathrm{MM}}=\frac{\mathrm{d}I_{\mathrm{M}}}{\mathrm{d}V_{\mathrm{M}}}$ is obtained by measuring the response of \var{I}{M} to \var{V}{M}.
Small offsets in measured conductances arise using the lock-in amplifiers, due to capacitances to ground within the electronics.
These offsets are calibrated using Coulomb blockaded measurements and corrected. Voltage-divider effects are not corrected, since we focus on low tunneling regimes (G $\ll$ 2e$^2$/h) where the device resistance is large compared to the resistances of the fridge lines and the current meters, such that the multi-terminal effect is small.

\subsection{RF-reflectometry measurements}
For fast characterisation of the device, we employ radio-frequency (RF) lead reflectometry~\cite{Vigneau2023} in addition to the DC/AC conductance measurements. 
Each ohmic contact is connected to an inductor, designed with varying inductances $L_{\mathrm{L, M, R}} = 0.2, 0.5, 1.5$~\SI{}{\micro \henry}, that together with a parasitic capacitance to ground via bond-wires result in resonators with frequencies of $f_{\mathrm{L, M, R}} = 723, 505, 248$~\SI{}{\mega\hertz}.
The complete circuit diagram including the fridge wiring and filters is presented in Ref.~\cite{Kulesh2024}.
Using a directional coupler, we obtain the reflected signal of each lead.
We denote with $S_{21}^{\mathrm{L}}$,  $S_{21}^{\mathrm{M}}$ and  $S_{21}^{\mathrm{R}}$ the normalised reflected signals of the left, middle and right lead respectively, which correspond roughly linearly to conductance.
All three signals can be measured simultaneously through multiplexing~\cite{Hornibrook2014}, using the circuit shown in \cref{fig:supp1}. 
The amplitude and phase of the complex reflected signal are translated into a single quantity $\widetilde S_{21}^{\mathrm{i}}$ and normalized, following Ref.~\cite{Kulesh2024}.
In combination with saw-tooth pulses on the QD plunger gates, generated by arbitrary waveform generators, this allows for scanning the parameter space many times faster than through DC measurements.

\subsection{Measurement procedures}
Magnetic fields are applied using a 3D vector magnet.
The field perpendicular to the superconducting loop (\var{B}{z}) is generated using a high-resolution current source, giving a \var{B}{z} resolution below \SI{0.1}{\micro\tesla} (providing sufficient resolution for the flux period of \SI{28}{\micro\tesla}). 
A small (but significant) hysteresis on the order of \SI{5}{\micro\tesla} is observed when sweeping \var{B}{z} in opposite directions.
This is counteracted by setting \var{B}{z} first to \SI{-20}{\micro\tesla} and then sweeping this field back in the positive direction, such that consecutive experiments where \var{B}{z} is varied are consistent.
To spin-polarise the QDs, a magnetic field of \SI{150} to \SI{200}{\milli\tesla} is applied parallel to the channel (\var{B}{x}).
Due to an imperfection in the alignment, this introduces a small \var{B}{z} component as well, on the order of 80 flux quanta. 
It was not possible to accurately correct this offset for this work, and so we are unable to determine the \var{B}{x} value that corresponds to precisely $0$ flux through the loop. To accurately determine this point of zero flux in future work, we recommend significantly smaller loop sizes.\\ \\
In the following sections, we briefly outline the details and procedures used to acquire the results in \cref{fig:Fig2} and \cref{fig:Fig3}. 
\subsubsection{Spin-flip procedure}
First, the tune-up procedure described in the main text is followed to ensure $\lvert{t_1}\rvert=\lvert\Delta_1\rvert$,$\lvert{t_2}\rvert=\lvert\Delta_2\rvert$ and all $\mu_i = 0$. Next, the flux dependent energy spectrum and zero-bias conductance in \cref{fig:Fig2}a and b are measured sequentially. The full energy spectrum is measured using RF-techniques, to allow for faster data collection, while the zero-bias features are specifically recorded through slower conductance measurements, ensuring more accurate phase extraction from fitting. In-between these two measurements, the flux-controlling magnetic field $B_{\mathrm{z}}$ is set to \SI{-20}{\micro\tesla} to mitigate any hysteresis effects. Next, we flip the spin on, for example, the left QD by changing its chemical potential energy. A slight adjustment is then made to the energy of the ABS that couples to the left QD, re-ensuring the $\lvert{t_1}\rvert=\lvert\Delta_1\rvert$ condition. Appropriate charge stability diagrams are collected to confirm that we are back at the sweet spot. We ensure that this adjustment to the ABS energy is small compared to its full extent. We now repeat the flux dependent spectrum measurement, allowing us to identify the spin-flip induced phase shift.

\subsubsection{ABS shift procedure}
To obtain the dataset in \cref{fig:Fig3}c, a sequence of steps is taken to correct for cross capacitances and renormalization effects.
The aim is to study the effective phase-shift experienced by the system when changing \ABS{2}.
After identifying a range where \ABS{2} modulates the energy of an ABS,  we set \ABS{2} to a starting point of \SI{-30}{\milli\volt}.
Here the effective couplings are $\lvert{t_1}\rvert=\lvert\Delta_1\rvert$, but $\lvert{t_2}\rvert\neq \lvert\Delta_2\rvert$.
We then take the following sequence of steps:
\begin{enumerate}
    \item Tune \var{V}{QDR} to select the down-spin resonance on QDR.
    \item Record Charge Stability Diagrams (CSDs) for both QDL-QDM and QDM-QDR, to keep track of the dominant interdot coupling and identify where the two-site sweet spots occur (see \cref{fig:supp4}).
    \item  Recalibrate all \var{V}{QDi} to their respective $\mu_i = 0$ points.
    \item Set the magnetic field $B_{\mathrm{z}}$ (controlling the flux) to \SI{-20}{\micro\tesla}, to mitigate any hysteresis effects.
    \item Measure the zero-bias conductance on the middle QD (\var{G}{MM}) as a function of the external flux.
    \item Reset $B_{\mathrm{z}}$ to \SI{-20}{\micro\tesla} and measure the finite-bias energy spectrum ($\tilde{S}^{\mathrm{M}}_{21}$) as a function of $B_{\mathrm{z}}$.
    \item Retune \var{V}{QDR} to select the up-spin resonance and repeat steps (2)-(7).
    \item Increase \ABS{2} by \SI{2}{\milli\volt}, to adjust the ABS charge and energy.
    \item Repeat steps (1)-(8) until the end of the chosen \ABS{2} range is reached.
\end{enumerate} 

\subsection{Theoretical results}
The calculations in \cref{fig:Fig1}c are reproduced from the results presented in Ref.~\cite{CX_scaling}, which extends the model from Refs.~\cite{Dominguez2016, Tsintzis2022}.
To model the device, consisting of 3 quantum dots and 2 hybrid segments, a five-site model is considered in a many-body Fock basis where charging energy and spin are included on each site.
In the perturbative limit with large charging and Zeeman energies, this model maps exactly to the 3-site Kitaev chain~\cite{luna2025}.
We refer to the aforementioned references for details.
To reproduce the experimental data in \cref{fig:Fig3}c, we extend the model by introducing an effective variation in the spin-orbit field. 
Below we introduce the general model used and derive \cref{eq:phasedependence} as presented in the main text.

\subsubsection{Phase shifts with a patially varying spin-orbit field}
A general approximation in modelling the system is that the spin-orbit field (along $\hat{y}$) is perpendicular to the chain and the applied Zeeman field (along $\hat{z}$).
Here, we describe the complex phase that arises when this constrained is relaxed. 
In the experimental system, the charging energy of the QDs is typically the largest energy scale in the system.
In order to simplify the discussion, we neglect charging energy and work in the BdG representation~\cite{luna2025,Luethi2025}.
We consider the subsystem consisting of three tunnel coupled sites:
\begin{equation}
    H = H^1_{\mathrm{QD}} +H^{1,2}_{\mathrm{T}} +H^2_{\mathrm{QD}} + H^{2,3}_{\mathrm{T}} + H^3_{\mathrm{QD}}
\end{equation}
where sites 1,3 include a Zeeman energy $E^{\mathrm{QD}}_{Z}$ and a chemical potential $\mu^i_{\mathrm{QD}}$, written:
\begin{equation}
    H^i_{\mathrm{QD}} = \psi_i^{\dagger}\left[\mu^{i}_{\mathrm{QD}}\tau_z\sigma_0+E^{\mathrm{QD}}_{\mathrm{Z}}\tau_0\sigma_z\right]\psi_i
\end{equation}
with $\psi_i$~=~($c_{i,\uparrow}$, $c_{i,\downarrow}$, $c^{\dagger}_{i,\downarrow}$, $-c^{i,\dagger}_{i,\uparrow}$) the subspace basis for site $i$, $\tau_i$ the Pauli matrices for particle-hole space and $\sigma_i$ the Pauli matrices for spin. 
The middle site has in addition an induced superconducting pairing $\Delta$ and a generally lower Zeeman energy $E^{\mathrm{ABS}}_{Z}$:
\begin{equation}
\label{eq:H_ABS_1}
    H^2_{\mathrm{QD}} = H_{\mathrm{ABS}}=\psi_i^{\dagger}\left[\mu_{\mathrm{ABS}}\tau_z\sigma_0+E^{\mathrm{ABS}}_{\mathrm{Z}}\tau_0\sigma_z+\Delta \tau_x\sigma_0\right]\psi_i
\end{equation}
To couple neighbouring sites $i$,$j$, we implement the most general hopping matrix (respecting time reversal symmetry) with normal hopping $t$ and spin-rotating hopping $t_{\mathrm{SO}}$ around a  unit vector $\vec{n}_{i,j}$. 
The spin precession angle $\theta_{ij}$ is set by $\tan{\theta_{ij}=\frac{t}{t_{\mathrm{SO}}}}$.
The tunneling Hamiltonian is then given by:
\begin{equation}
\label{eq:HT_general}
H^{i,j}_{T}=\psi_i\left[t\tau_z\sigma_x+t_{\mathrm{SO}}\tau_z\vec{\sigma}\cdot\vec{n}_{i} \right]\psi_j
\end{equation}
where $\vec{\sigma}=[\sigma_x,\sigma_y,\sigma_z]$. 
Typically $\vec{n}$ is set to $[0,1,0]$, reflecting a purely 1D Rashba spin-orbit term. 
Here we consider the general case  $\vec{n}=[n_x,n_y,n_z]$. Lastly, we set $\mu^i_{\mathrm{QD}}=E^{\mathrm{QD}}_{\mathrm{Z}}$, assuming the QDs are weakly coupled to the ABS~\cite{Liu2024_enhancedgap}.
We are interested in the complex phase that remains when mapping this spinful three-site system to a spinless two site system. 
To derive this, we use Pymablock to perform a Schrieffer-Wolff transformation~\cite{Pymablock_article, Pymablock_article_codebase} and project out the middle site and the undesired spin.
This leaves an effective ECT coupling $t_{\mathrm{eff}}$ and effective CAR coupling $\Delta_{\mathrm{eff}}$ between the two sites. 
We can write these coupling terms as:
\begin{equation}
    \begin{aligned}
        t_{\mathrm{eff}}&=\frac{c_1\mu_{\mathrm{ABS}} +c_2E^{\mathrm{ABS}}_{\mathrm{Z}}}{(E^{\mathrm{ABS}}_{\mathrm{Z}})^2-\Delta^2-(\mu_{\mathrm{ABS}})^2}\\
        \Delta_{\mathrm{eff}}&=\frac{\Delta t_{\mathrm{SO}}c_3}{(E^{\mathrm{ABS}}_{\mathrm{Z}})^2-\Delta^2-(\mu_{\mathrm{ABS}})^2}
    \end{aligned}
\end{equation}
where the complex coefficients $c_1$,$c_2$,$c_3$ are given by:
\begin{equation}
    \begin{aligned}
c_1&=t^2+t_{\mathrm{SO}}^2\left(\vec{n}_{1}\cdot\vec{n}_{2}-2(\vec{n}_{1}\cdot\vec{n}_{2})_{\hat{z}}\right)-i\left[tt_{\mathrm{SO}}(\vec{n}_1+\vec{n}_2)_{\hat{z}}+t_{\mathrm{SO}}^2(\vec{n}_1\times \vec{n}_2)_{\hat{z}}\right]\\
        c_2&=t_{\mathrm{SO}}^2\vec{n}_1\cdot\vec{n}_2-t^2+i\left[tt_{\mathrm{SO}}(\vec{n}_1+\vec{n}_2)_{\hat{z}}-t_{\mathrm{SO}}^2(\vec{n}_1\times \vec{n}_2)_{\hat{z}}\right]\\
        c_3&= [t_{so}(\vec{n}_1\times\vec{n_2})_{\hat{y}}-t(\vec{n}_{1}+\vec{n}_{2})_{\hat{y}}]+i[t(\vec{n}_1+\vec{n}_2)_{\hat{x}}-t_{\mathrm{SO}}(\vec{n}_1\times\vec{n_2}_{\hat{x}})]
    \end{aligned}
\end{equation}

Since we are only interested in the change in complex phase as a function of $\mu_{\mathrm{ABS}}$, we can neglect $\Delta_{\mathrm{eff}}$ as it only contributes a constant offset.
This leaves the phase contribution from $t_{\mathrm{eff}}$, which we can write in the form:
\begin{equation}
\label{eq:arg_teff}
    \arg(t_{\mathrm{eff}}) = \arg(c_{1})+\arg{\left(\mu_{\mathrm{ABS}}+E^{\mathrm{ABS}}_{Z}\Re\left(\frac{c_2}{c_1}\right)+iE^{\mathrm{ABS}}_{Z}\Im\left(\frac{c_2}{c_1}\right)\right)}
\end{equation}
From here, we extract the change in phase $\widetilde{\phi}$ as a function of $\mu_{\mathrm{ABS}}$, where we only need to consider the right term in \cref{eq:arg_teff}:
\begin{equation}
\label{eq:general_result}
    \widetilde{\phi}(\mu_{\mathrm{ABS}})=\frac{\pi}{2}-\arctan\left(\frac{\mu_{\mathrm{ABS}}+E^{\mathrm{ABS}}_{\mathrm{Z}}\Re(\frac{c_1}{c_2})}{E^{\mathrm{ABS}}_{\mathrm{Z}}\Im{\left(\frac{c_1}{c_2}\right)}}\right)+\phi_0
\end{equation}
where $\phi_0$ is the constant offset added by the phase contribution of $\Delta_{\mathrm{eff}}$ and $\arg(c_1)$.
This result is the general form of the result \cref{eq:phasedependence} presented in the main text.
\subsubsection{Gauge transformation from spin-orbit field to Zeeman field}
The above result relies on the physically reasonable assumption that a realistic device may have a general spatially varying spin orbit-field that is not necessarily strictly perpendicular to the chain.
To simplify the discussion in the main text, we present a simpler case where the spin-orbit field is fixed to be perpendicular and instead the Zeeman field in the ABS is rotated in the y-z plane, captured by a single angle $\theta_{\mathrm{A}}$.
We show here the equivalence between these two cases via a unitary transformation, where the latter matches a specific orientation of a spatially varying spin-orbit field.
With the magnetic field in $H_{\mathrm{ABS}}$ misaligned in the z-y plane with an angle $\theta_{\mathrm{A}}$ we have the following Hamiltonian for the middle site:
\begin{equation}
\label{eq:H_ABS_2}
    H^{\theta}_{\mathrm{ABS}} = \psi_i^{\dagger}\left[\mu_{\mathrm{ABS}}\tau_z\sigma_0+\Delta \tau_x\sigma_0+E^{\mathrm{ABS}}_{\mathrm{Z}}\cos(\theta_{\mathrm{A}})\tau_0\sigma_z+E^{\mathrm{ABS}}_{\mathrm{Z}}\sin(\theta_{\mathrm{A}})\tau_0\sigma_y\right]\psi_i
\end{equation}
while the tunneling Hamiltonians, fixing $\vec{n}=[0,1,0]$, are:
\begin{equation}
\label{eq:HT_specific}
    H^{i,j}_T =\psi_i(t\tau_z\sigma_x+it_{\mathrm{SO}}\tau_z\sigma_y)\psi_j
\end{equation}
with $t=t_0\cos{\theta_{\mathrm{SO}}}$ and  $t_{\mathrm{SO}}=t_0\sin{\theta_{\mathrm{SO}}}$.
We can express these matrices as $T=e^{i\theta_{\mathrm{SO}}\sigma_y}$.
Utilizing the unitary transformation $U$ that diagonalizes $H^{\theta}_{\mathrm{ABS}}$ (\cref{eq:H_ABS_2}) we can transform the tunneling Hamiltonians $H^{1,2}_{T}$ and $H^{2,3}_{T}$ using:
\begin{equation}
    TU=e^{i\theta_{\mathrm{SO}}\sigma_{y}}e^{i\frac{\theta_A}{2}\sigma_{x}}\qquad\qquad U^{\dagger}T=e^{-i\frac{\theta_A}{2}\sigma_x}e^{i\theta_{\mathrm{SO}}\sigma_y}
\end{equation}
To match this to the general tunneling Hamiltonian \cref{eq:HT_general}, we then need to solve:
\begin{equation}
    e^{{i\theta_{12}}\vec{\sigma}\cdot\vec{n}_{1,2}}=e^{i\theta_{\mathrm{SO}}\sigma_{y}}e^{i\frac{\theta_A}{2}\sigma_{x}}\qquad\qquad e^{i\theta_{23}\vec{\sigma}\cdot\vec{n}_{2,3}}=e^{-i\frac{\theta_A}{2}\sigma_x}e^{i\theta_{\mathrm{SO}}\sigma_y}
\end{equation}
which yields the solution:
\begin{equation}
\begin{aligned}
    \vec{n}_1&=\frac{1}{\sin(\theta_{12})}\left[\cos(\theta_{\mathrm{SO}})\sin(\theta_{\mathrm{A}}/2), ~\sin(\theta_{\mathrm{SO}})\cos{(\theta_{\mathrm{A}}/2)},~\sin(\theta_{\mathrm{SO}})\sin(\theta_A/2) \right]\\
    \vec{n}_2&=\frac{1}{\sin(\theta_{23})}\left[-\cos(\theta_{\mathrm{SO}})\sin(\theta_{\mathrm{A}}/2), ~\sin(\theta_{\mathrm{SO}})\cos{(\theta_{\mathrm{A}}/2)},~\sin(\theta_{\mathrm{SO}})\sin(\theta_{\mathrm{A}}/2) \right]
\end{aligned}
\end{equation}
and $\theta_{12}=\theta_{23}$. 
The Zeeman field alignment in the ABS can thus be interpreted as an effective spin-orbit field variation, where the spin-precession is equal  but around two vectors pointing symmetrically in different vectors.
While this is physically less plausible as underlying mechanism, it captures the essence of the experimental result. 

\subsubsection{Phase shift with offset in QD-ABS polarization axis}
We now repeat the extraction of the complex phase in the system as a function of $\mu_{\mathrm{ABS}}$, with the simpler Hamiltonian terms \cref{eq:H_ABS_2} and \cref{eq:HT_specific}.
Again performing a Schrieffer-Wolff transformation, we obtain new expressions for the effective couplings $\Delta_{\mathrm{eff}}$ and $t_{\mathrm{eff}}$:
\begin{equation}
    \Delta_{\mathrm{eff}}=\frac{2\Delta tt_{\mathrm{SO}}}{(E^{\mathrm{ABS}}_{\mathrm{Z}})^2-\Delta^2-(\mu_{\mathrm{ABS}})^2}
\end{equation}
and:
\begin{equation}
    t_{\mathrm{eff}}=\frac{E^{\mathrm{ABS}}_{\mathrm{Z}}(t^2+t^2_{\mathrm{SO}})\cos(\theta_{\mathrm{A}})+(t^2-t^2_{\mathrm{SO}})\mu_{\mathrm{ABS}}}{(E^{\mathrm{ABS}}_{\mathrm{Z}})^2-(\Delta)^2-(\mu
    _{\mathrm{ABS}})^2}+i\frac{2E^{\mathrm{ABS}}_{\mathrm{Z}}tt_{\mathrm{SO}}\sin(\theta_{\mathrm{A}})}{(E^{\mathrm{ABS}}_{\mathrm{Z}})^2-(\Delta)^2-(\mu_{\mathrm{ABS}})^2}
\end{equation}
Since the expression for CAR is real, the remaining phase in the system is given by:
\begin{equation}
\label{eq:main_result}
    \widetilde{\phi}(\mu_{\mathrm{ABS}})=\arg(t_{\mathrm{eff}})=\arctan\left(\frac{\Im(t_{\mathrm{eff}})}{\Re(t_{\mathrm{eff}})}\right)=\frac{\pi}{2}-\arctan\left(\frac{\mu_{\mathrm{ABS}}+\mu_0}{\Gamma}\right)
\end{equation}
with $\Gamma=2E^{\mathrm{ABS}}_z\frac{tt_{\mathrm{SO}}}{t^2-t_{\mathrm{SO}}^2}\sin\left(\theta_{\mathrm{A}}\right)$ and $\mu_0=E^{\mathrm{ABS}}_{\mathrm{z}}\frac{t^2+t_{so}^2}{t^2-t_{so}^2}\cos(\theta_{\mathrm{A}})$ as shown in main text.
We again stress that this result only predicts a total change in $\widetilde\phi$ of $\pi$ when sweeping $\mu_{\mathrm{ABS}}$.
In order to fit the equation in \cref{fig:supp5}, we add a fitting parameter $\delta\widetilde\phi$ which sets the separation between the two phase plateaus.

\subsubsection{Alternative considerations}
The above result, considering a spatial variation in the spin-orbit field, can be a plausible explanation for the experimental result.
We stress, however, that further experimental work is needed to pinpoint the exact mechanism, and that alternative effects may need to be considered.
A similar connection between phase and charge arises, for example, in electron scattering processes involving a screened impurity in a metal. 
In this context, the \textit{Friedel sum rule} connects the phase $\tilde{\phi}$ gained by an incoming electron to the charge of the impurity $\langle \hat{N}\rangle$ via~\cite{Langreth1966}:
\begin{equation}
    \tilde{\phi}= \pi\langle \hat{N}\rangle
\end{equation}
Since the ABS in our system is essentially a spin-impurity strongly coupled to a superconductor, as considered in the typical YSR-state picture~\cite{Rasmussen2018}, we posit that similar effects may arise when the excited state of the system is a screened doublet.
A similar consideration was recently discussed in Ref.~\cite{Hashimoto2024}.
In the superconducting atomic limit, the charge of the ABS can be readily extracted as a function of $\mu_{\mathrm{ABS}}$.
This would give the following relation between phase and $\mu_{\mathrm{ABS}}$:
\begin{equation}
\label{eq:friedel}
    \widetilde\phi=\frac{2|v|^2}{\pi}=\pi\left(1-\frac{\mu_{\mathrm{ABS}}}{\sqrt{\mu_{\mathrm{ABS}}^2+\Delta^2}}\right)
\end{equation}
Notably, since the total, smooth increase in charge across an ABS in the singlet ground state is $2e$, the phase shift induced by this mechanism would in fact be more than $\pi$ (namely 2$\pi$). 
Nevertheless, this does not account for the seemingly arbitrary separation observed in \cref{fig:Fig3}. 
We include it here as a speculative comparison, fitted against the experimental data in \cref{fig:supp5}.
We further note that smooth phase evolutions as a function of chemical potential are well-known in transport through a single quantum dot in an interference loop, where a similar discussion arose~\cite{Schuster1997,Yacoby1995,LevyYeyati1995}.
\newpage

\subsection{Device characterization}
\begin{figure}[h]
    \centering
    \includegraphics[width=\linewidth]{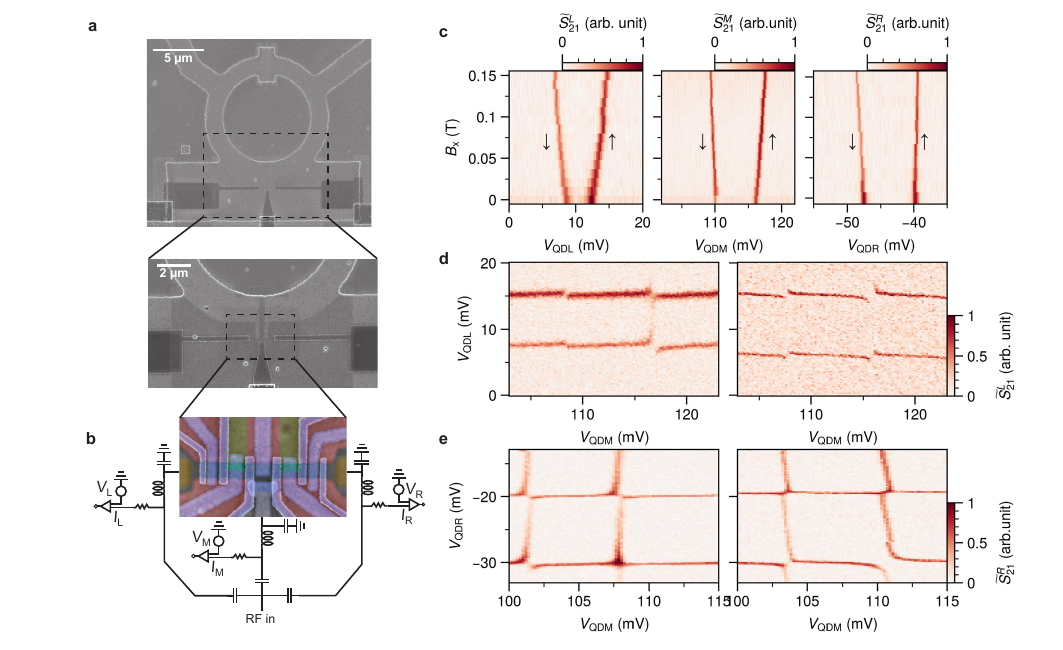}
    \caption{\textbf{Device and characterization} \textbf{a)} Scanning electron microscopy (SEM) images of the measured device, showing the full flux loop (top) and the part of the loop that connects to the device region (bottom). Light areas indicate remaining Aluminium, while the three darker strips are the Ti/Pd device leads 
    \textbf{b)} Zoomed in false-coloured SEM of the active region after all three gate depositions. Off-chip lumped element resonators are connected to all three normal leads, allowing for fast reflectometry measurements. 
    \textbf{c)} Zeeman splitting of all three QDs used for the measurements of \cref{fig:Fig2}, showing the energy splitting of the two spin states. 
    \textbf{d)} Charge stability diagrams of the left and middle QD, showing both CAR (left) and ECT (right) dominated coupling, indicating the existence of sweet spots for both spin states on the left QD. 
    \textbf{e)} Charge stability diagrams of the middle and right QD, showing similar behaviour.}
    \label{fig:supp1}
\end{figure}

\clearpage

\subsection{Reproducibility}

\begin{figure}[h]
    \centering
    \includegraphics[width=\linewidth]{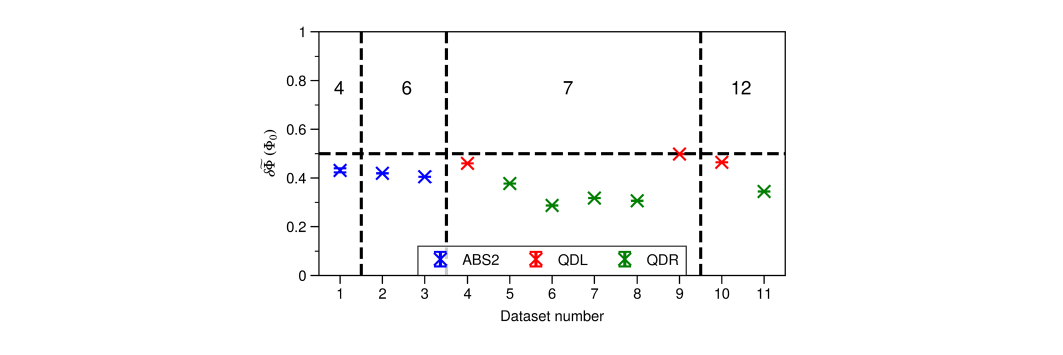}
    \caption{\textbf{Compilation of datasets demonstrating phase shifts reproducibility.} The flux difference $\delta \widetilde{\Phi}(\Phi_0)$ is defined as the flux shift between two differently tuned 3-site chains, where the flux $\widetilde{\Phi}$ of each is extracted using the procedure shown in \cref{fig:supp3}. 
    Different colours indicate which element of the system is responsible for the phase shift. 
    Numbers in the top half of the plot specify the number of times the device was warmed up to room temperature and cooled down again before the data was measured. 
    All points are projected onto $\delta \widetilde \Phi < \frac{\Phi_0}{2}$, since we are unable to distinguish between shifts larger or smaller than $\pi$ from the relative shifts only. The horizontal dashed line indicates and ideal $\pi$-shift.}
    \label{fig:supp2}
\end{figure}

\subsection{Fitting procedure}
\begin{figure}[h]
    \centering
    \includegraphics[width=\linewidth]{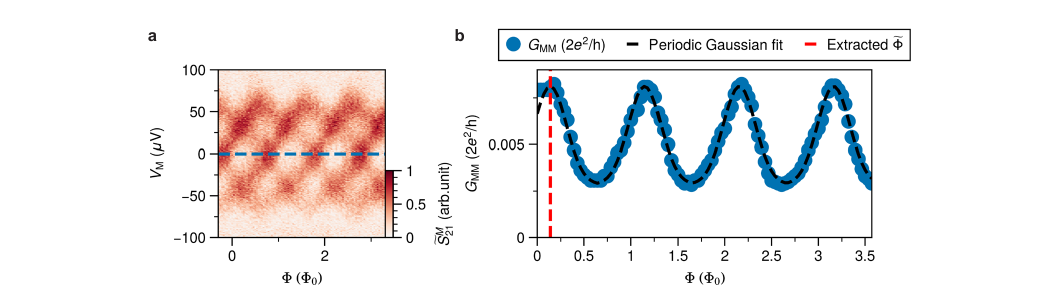}
    \caption{\textbf{Procedure of extracting $\widetilde\Phi$.} \textbf{a)} An example of a flux dependent energy spectrum of a tuned up 3-site chain. \textbf{b)} The flux dependent zero bias conductance is fitted to a periodically repeated Gaussian of the form $A_0 +  Ae^{-\frac{(\Phi^*-\frac{\Phi_0}{2})^2}{2\sigma^2}}$, where $\Phi^*= (\Phi -\widetilde\Phi + \frac{\Phi_0}{2})$ (mod $\Phi_0$). Fitting this function allows for automatic extraction of the flux offset. 
    The extracted flux $\widetilde \Phi$ for this example is indicated by the red dashed line. 
    For some of the data in \cref{fig:supp2}, this extraction is performed on the reflectometry signal $S_{21}^{\mathrm{M}}$, 
    rather than the conductance \var{G}{MM}. There, it was found that inverted Gaussians ($A<0$) were a better fit, likely due to the non-linear relation between $S_{21}^{\mathrm{M}}$ and \var{G}{MM}.}
    \label{fig:supp3}
\end{figure}

\clearpage

\subsection{Sweet spot determination along the ABS}

\begin{figure}[h]
    \centering
    \includegraphics[width=\linewidth]{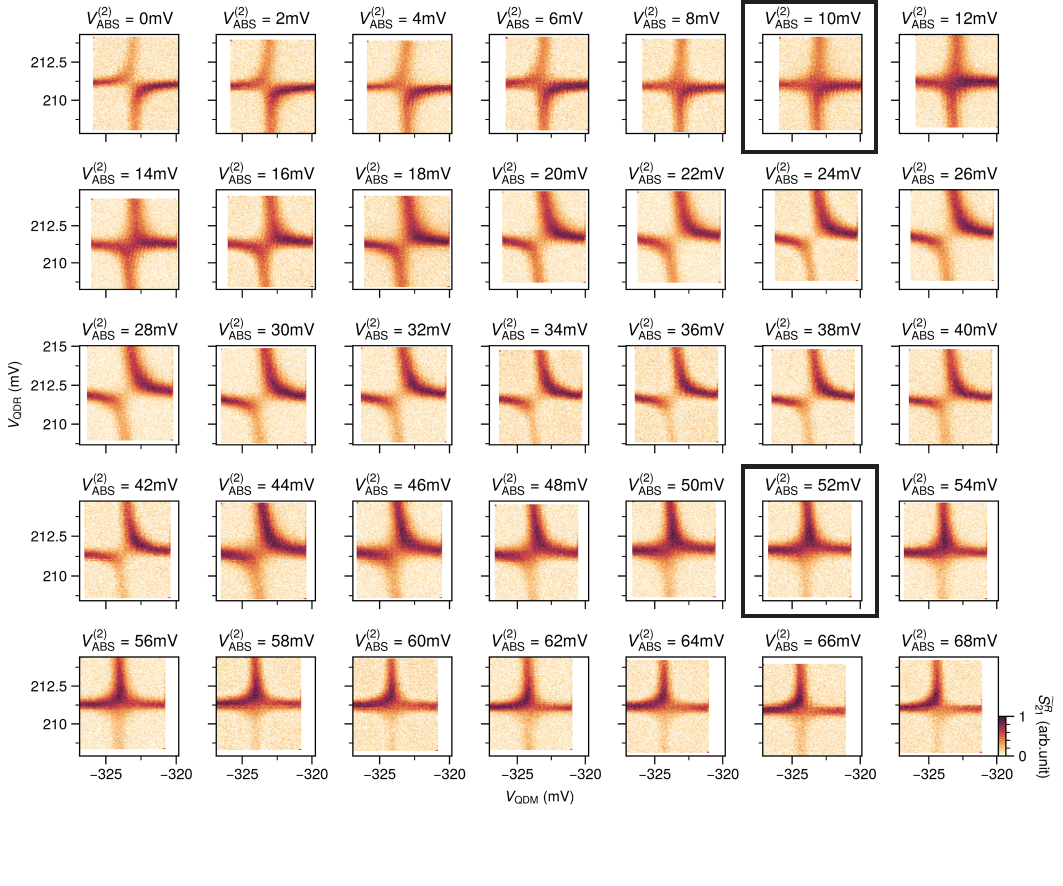}
    \caption{\textbf{Interaction between the middle QD and the up-spin of the right QD for varying ABS energy.} Full set of CSDs used to identify relevant regimes. Starting at $V_{\mathrm{ABS}}^{(2)} =0$, we observe transitions from $\Delta>t$ to $\Delta<t$ and back to $\Delta>t$, indicating the existence of two sweet spots at $V_{\mathrm{ABS}}^{(2)}\approx $ \SI{10}{\milli\volt} and $V_{\mathrm{ABS}}^{(2)}\approx $ \SI{52}{\milli\volt}, highlighted by the black squares.}
    \label{fig:supp4}
\end{figure}

\clearpage

\subsection{Zero bias flux dependence along the ABS}

\begin{figure}[h]
    \centering
    \includegraphics[width=\linewidth]{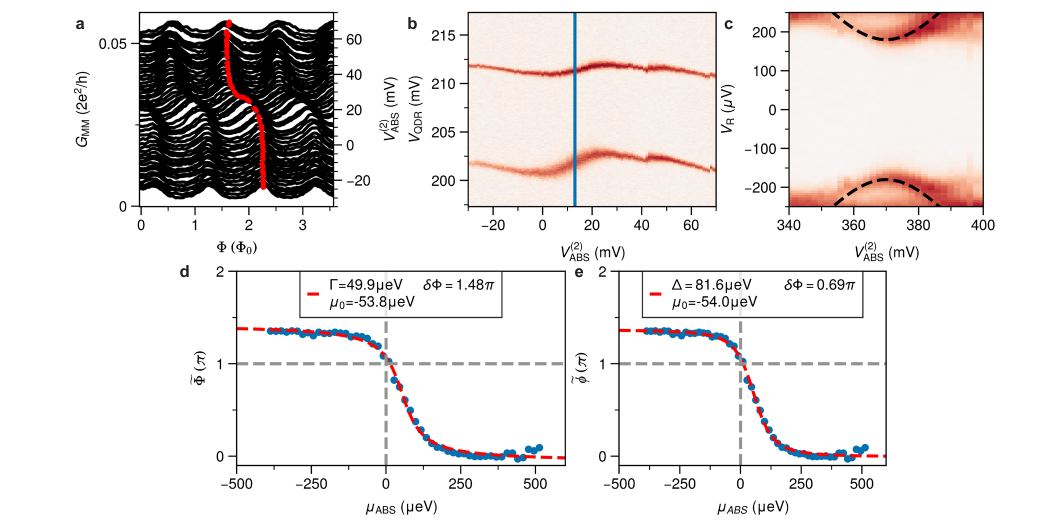}
    \caption{\textbf{Fitting the flux-dependent conductance measurements as a function of \ABS{2}}. \textbf{a)} Compilation of the raw measurements that are used to extract the flux-shift $\widetilde{\Phi}$ presented in \cref{fig:Fig3}c, following the measurement procedure described in Methods.
    Every linetrace is offset by 0.002 $\frac{e^2}{h}$.
    The selected peaks used to extract $\widetilde{\Phi}$ in \cref{fig:Fig3}c are overlaid.
    From the continuous measurement it is clear that the total shift exceeds $\Phi_0/2$.
    To fit the extracted data to the theoretical results, we first map \ABS{2} to chemical potential $\mu_{\mathrm{ABS}}$.
    \textbf{b)} From the charge stability diagram between \var{V}{QDR} and \ABS{2} (\cref{fig:Fig3}a) we can determine the value \ABS{2} corresponding to $\mu_{\mathrm{ABS}}=0$.
    \textbf{c)} Tunneling spectroscopy allows to estimate the leverarm to convert \ABS{2} to $\mu_{\mathrm{ABS}}$.
    \textbf{d)} 
    Fitting of the experimentally observed flux-shift $\widetilde{\Phi}$ as a function of $\mu_{\mathrm{ABS}}$, to the derived \Cref{eq:phasedependence}. 
    Since the experimental data has two plateaus separated by more than $\pi$, an arbitrary fitting parameter $\delta\Phi$ needed to be included that captures this effect (see Methods). 
    \textbf{e)} Similarly, we fit the experimental data to the Friedel phase shift (\Cref{eq:friedel}), again with an arbitrary scaling parameter $\delta\Phi$.
    We note that this equation similarly described the data reasonably well, but the validity is speculative.
    }
    \label{fig:supp5}
\end{figure}

\clearpage 

\subsection{Flux dependent energy spectrum along the ABS for the up spin}
\begin{figure}[h]
    \centering
    \includegraphics[width=0.9\linewidth]{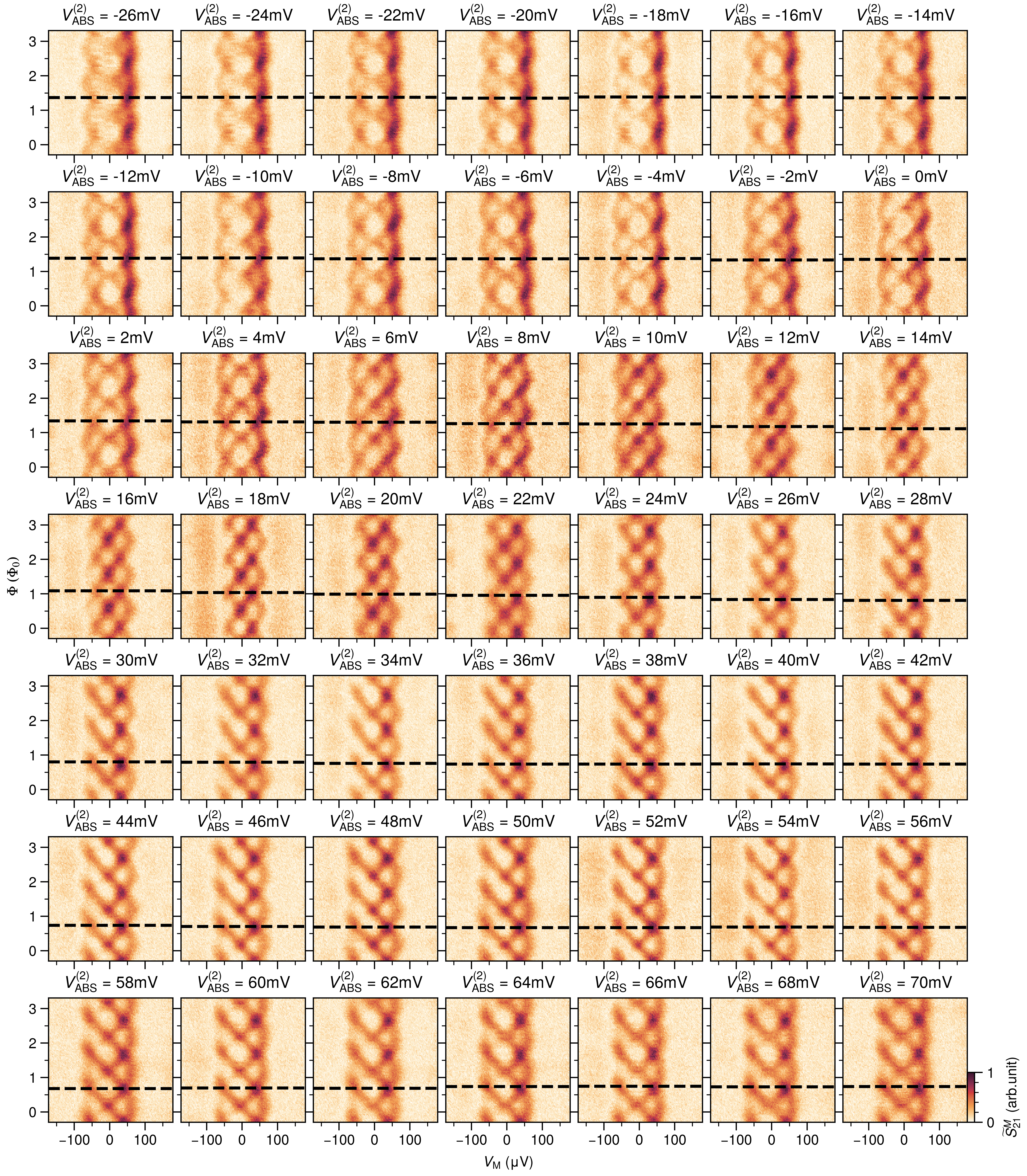}
    \caption{\textbf{Flux dependent energy spectrum of the up-spin resonance for different \ABS{2}}. Finite bias spectroscopy explicitly showing the smooth flux-shift of the up spin presented in \cref{fig:Fig3}c. Horizontal dashed lines indicate a flux where the gap is maximized.}
    \label{fig:supp6}
\end{figure}
\clearpage

\subsection{Flux dependent energy spectrum along the ABS for the down spin}
\begin{figure}[h]
    \centering
    \includegraphics[width=0.9\linewidth]{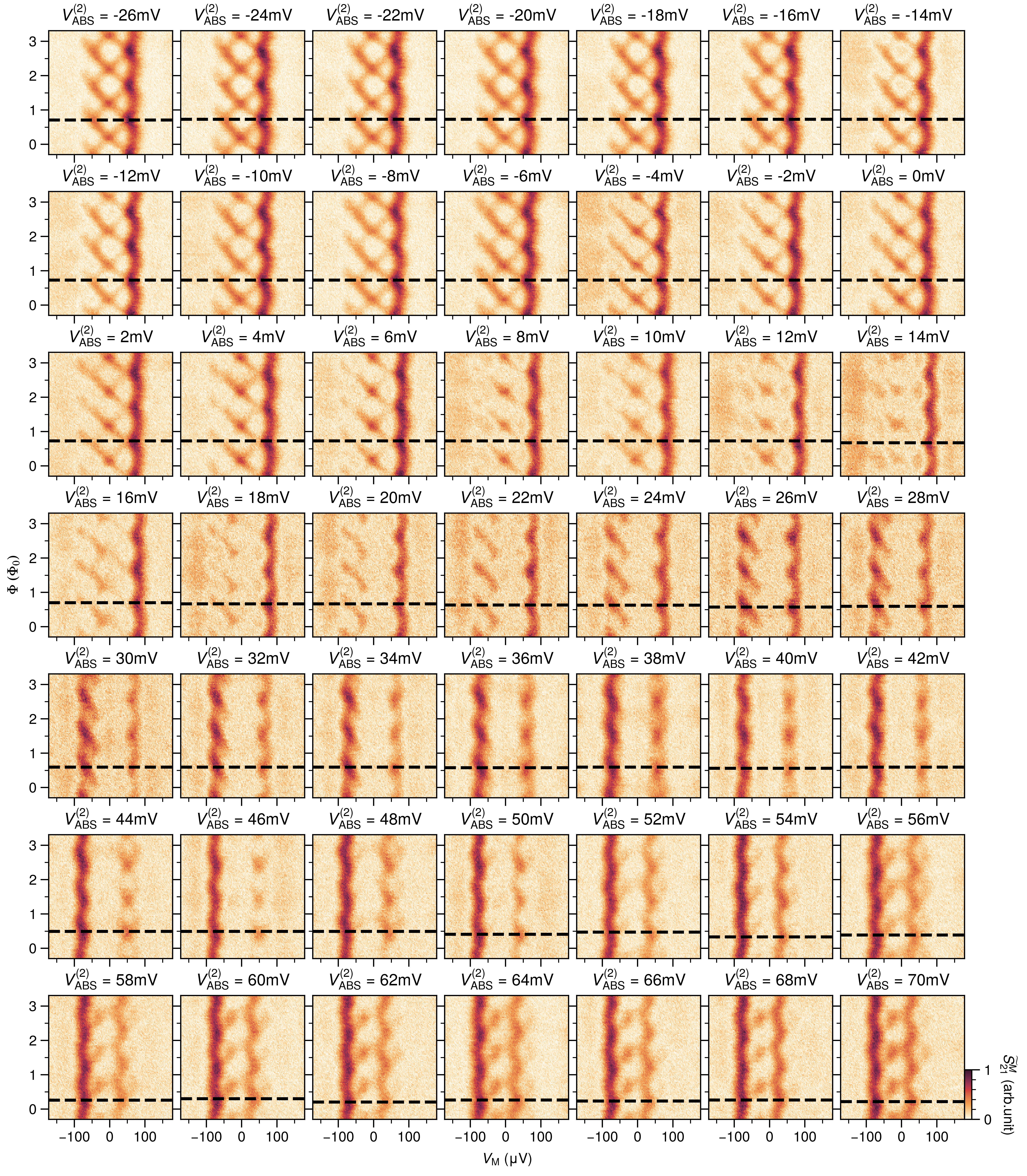}
    \caption{\textbf{Flux dependent energy spectrum of the down-spin resonance for different \ABS{2}}. Horizontal dashed lines indicate a flux where the gap is maximized. From this data it becomes clear why the extraction becomes unreliable for \ABS{2} $<$ \SI{16}{\milli\volt}, as the \var{S}{21} response becomes too weak around zero-bias to resolve any oscillations.}
    \label{fig:supp7}
\end{figure}

\clearpage

\subsection{Alternative phase extraction method}

\begin{figure}[h]
    \centering
    \includegraphics[width=\linewidth]{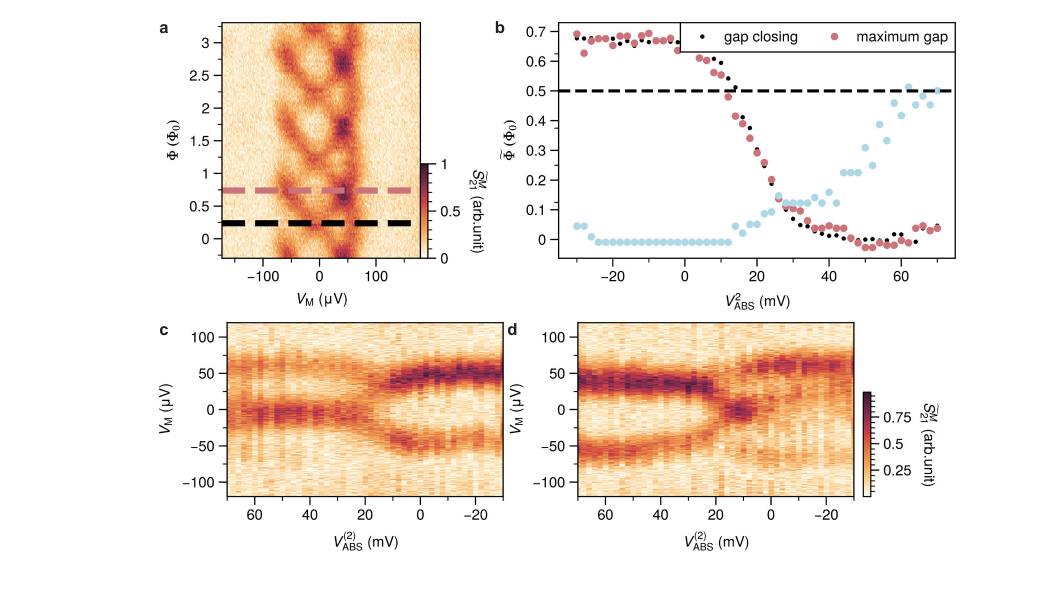}
    \caption{\textbf{Alternative phase extraction method} In the main text we chose the phase where the gap closes as reference point to track phase shifts in the system. 
    To ensure our result does not depend on the choice of reference point, we compare here with an alternative choice where the gap is maximal instead.
    We were unable to satisfyingly automate this process and instead utilise a manual extraction for this analysis.
    \textbf{a)} Finite bias spectroscopy at \ABS{2} = \SI{70}{\milli\volt}. 
    Points where the gap is either closed or maximal are highlighted as reference.
    \textbf{b)} Phase-shifts as a function of the ABS energy, comparing the two methods. 
    In both cases the evolution follows very similar trends and we find that the choice of extraction method does not affect the conclusions made in the main text.
    The manually selected phases are shown as dashed lines in \cref{fig:supp6} and \cref{fig:supp7}.
    To see the effect of the larger-than-pi shift, we extract $S_{21}$ linetraces at fixed flux $\Phi$ as a function of \ABS{2}. 
    \textbf{c)} and \textbf{d)} show these traces, starting at the black line and pink line in a) respectively. 
    In c), when the gap starts closed, the gap is reopened by tuning \ABS{2}.
    In d), the opposite occurs, where the gap is closed at \SI{10}{\milli\volt}
    Note that the gap reopens again, precisely because the induced phase-shift exceeds a $\pi$-shift.
    }
    \label{fig:supp8}
\end{figure}

\clearpage
\subsection{Zeeman dependent phase differences}
\begin{figure}[h]
    \centering
    \includegraphics[width=\linewidth]{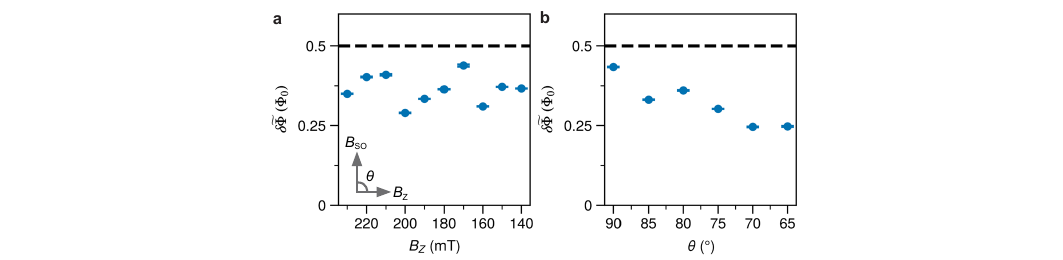}
    \caption{\textbf{Phase difference between two spin configurations as a function of the Zeeman field strength and orientation.} \textbf{a)} Varying Zeeman field at an angle close to perpendicular to the spin-orbit field ($\theta = 90$\textdegree). Significant variation in the extracted phase difference is attributed to the effect of \var{B}{Z} on the energy of the ABS. \textbf{b)} Varying the orientation $\theta$ of \var{B}{Z} = \SI{200}{\milli\tesla} with respect to the spin-orbit field. The two datasets measured in a) and b) are measured in electrostatically different configurations.}
    \label{fig:supp9}
\end{figure}

\section{Excitation gap reduction}
\begin{figure}[h]
    \centering
    \includegraphics[width=\linewidth]{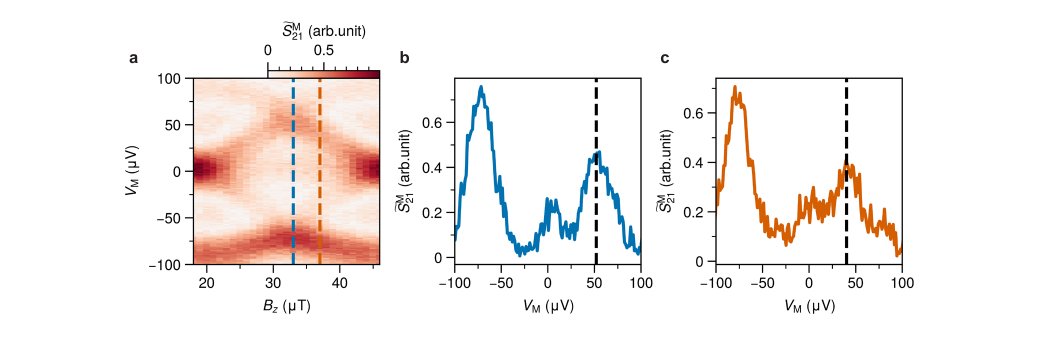}
    \caption{\textbf{Estimation of the excitation gap reduction resulting from the $\pi$ shift deviation.} \textbf{a)} Flux dependent spectroscopy of a 3-site chain. \textbf{b,c)} Extracted linecuts taken at the maximum energy gap and at a \SI{4}{\micro\tesla} deviation respectively. Using b) and c) we estimate a reduction in energy gap from \SI{50}{\micro\electronvolt} to \SI{38}{\micro\electronvolt}.}
    \label{fig:supp10}
\end{figure}
\end{widetext}
\end{document}